\documentstyle[12pt,world_sci,epsf]{article}
\begin{document}

\title{{\bf $B$ MESON LIFETIMES AT CDF}}
\author{JOHN E. SKARHA\thanks{For the CDF Collaboration;
Currently Guest Scientist at Fermi National Accelerator Laboratory,
P.O. Box 500, M.S. 318,
Batavia, IL 60510, USA}\\
Department of Physics and Astronomy\\
The Johns Hopkins University\\
Baltimore, Maryland 21218, USA}
\vspace{0.3cm}

\maketitle
\setlength{\baselineskip}{2.6ex}

\begin{center}
\parbox{13.0cm}
{\begin{center} ABSTRACT \end{center}
{\small \hspace*{0.3cm} Measurements of the $B_u$, $B_d$, and $B_s$ meson
lifetime using semileptonic $B_u \rightarrow e \nu D^0 X,
B_d \rightarrow e \nu D^* X, B_s \rightarrow l \nu D_s X$  events and exclusive
$B_u \rightarrow \psi^{(\prime)} K^{(*)},
B_d \rightarrow \psi^{(\prime)} K^{(*)}_{(s)}, B_s \rightarrow \psi \phi$
events are presented.  These results
used the precise position measurements of the CDF SVX silicon vertex detector
and were obtained from a 19.3 pb$^{-1}$ sample of 1.8 TeV $\overline{p}p$
collisions collected in 1992-93 at the Fermilab Tevatron collider.
Comparisons with previous measurements will be shown.}}
\end{center}

\section{Introduction}
During the 1992-93 Tevatron collider Run Ia, the Collider Detector at
Fermilab (CDF)~\cite{CDF} collected a data sample
of $\overline{p}p$ collisions at $\sqrt{s} = 1.8$ TeV with an integrated
luminosity of 19.3 pb$^{-1}$.  This
data sample, in combination
with improvements to the data acquisition system, the muon coverage, and
most importantly, the installation of the CDF SVX silicon vertex
detector~\cite{svx},
has allowed the first measurements of inclusive and exclusive $B$ meson
lifetimes at a hadron collider.
In this paper we report results on the $B_u$, $B_d$, and
$B_s$ meson lifetime using semileptonic $B_u \rightarrow e \nu D^0 X,
B_d \rightarrow e \nu D^* X, B_s \rightarrow l \nu D_s X$  events and exclusive
$B_u \rightarrow \psi^{(\prime)} K^{(*)},
B_d \rightarrow \psi^{(\prime)} K^{(*)}_{(s)}, B_s \rightarrow \psi \phi$
events.

\section{Charged and Neutral $B$ Meson Lifetimes}
Measuring the lifetime differences of the individual $B$ mesons is a direct
probe to possible non-spectator contributions in $B$ meson decay.
Only small lifetime differences are expected among the
different $B$ mesons (possibly as low as
$\sim 5\%$~\cite{thlife}) and experiments are now approaching this precision.

At CDF, the measurement of the charged and neutral $B$ meson lifetimes was
performed using fully reconstructed $B$ decays in the following
modes~\cite{charge}:
\begin{displaymath}
\begin{array}[b]{rllrll}
B^+ &\rightarrow J/\psi K^+       &\rightarrow \mu^+\mu^- K^+; &
B^+ &\rightarrow J/\psi K^{*+}    &\rightarrow \mu^+\mu^- K^0_s\pi^+ \\
B^+ &\rightarrow \psi (2S) K^+    &\rightarrow \mu^+\mu^-\pi^+\pi^- K^+; &
B^+ &\rightarrow \psi (2S) K^{*+} &\rightarrow \mu^+\mu^-\pi^+\pi^-
K^0_s\pi^+\\
B^0 &\rightarrow \psi K^0_s       &\rightarrow \mu^+\mu^- K^0_s; &
B^0 &\rightarrow \psi K^{*0}      &\rightarrow \mu^+\mu^- K^+\pi^- \\
B^0 &\rightarrow \psi (2S) K^0_s  &\rightarrow \mu^+\mu^-\pi^+\pi^- K^0_s; &
B^0 &\rightarrow \psi (2S) K^{*0} &\rightarrow \mu^+\mu^-\pi^+\pi^- K^+\pi^-
\end{array}
\end{displaymath}
These measurements using exclusive decay modes are rather unique, provide a
statistical precision of $10 - 12\%$, and are now published~\cite{cdf_exlife}.
We quote only the results here:
\begin{displaymath}
\begin{array}[b]{rll}
\tau^{+}_{exc} & = & 1.61\pm 0.16 {\rm (stat)}\pm 0.05 {\rm (syst) ps} \\
\tau^{0}_{exc} & = & 1.57\pm 0.18 {\rm (stat)}\pm 0.08 {\rm (syst) ps} \\
\tau^{+}_{exc}/\tau^{0}_{exc} & = & 1.02\pm 0.16 {\rm (stat)}\pm 0.05
{\rm (syst)}
\end{array}
\end{displaymath}
\vspace{-0.5in}
\begin{figure}[thb]
%\hspace{0.05in}
%\parbox[b]{3.2in}{\epsfxsize=3.2in\epsfbox{life_ed_d0_lxy1.ps}}%
\parbox[b]{3.0in}{\epsfxsize=3.0in\epsfbox{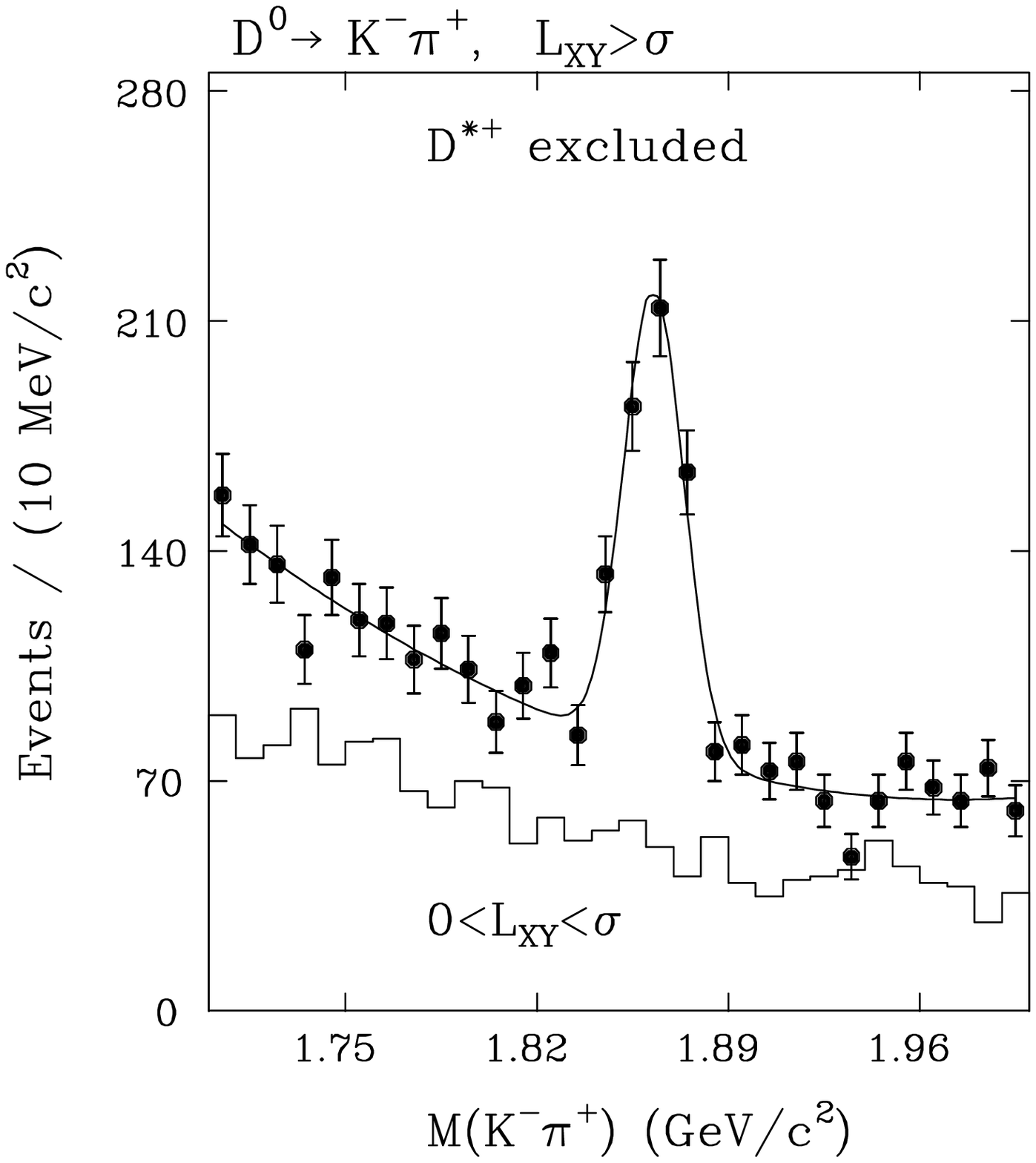}}%
\makebox[0in][r]{\raisebox{3.2in}{\em  a) \hspace{2.1in}}}
%\makebox[0in][r]{\raisebox{0.0in}{\em  a) \hspace{3.0in}}}
%\makebox[0in][r]{\raisebox{0.0in}{\em  a) $\; \; \; \; \; \; \; \;$}}
%\makebox[2.5in][l]{\em a)}
\hfill
%\parbox[b]{3.2in}{\epsfxsize=3.2in\epsfbox{LIFE_ED_DST.PS}}%
\parbox[b]{3.0in}{\epsfxsize=3.0in\epsfbox{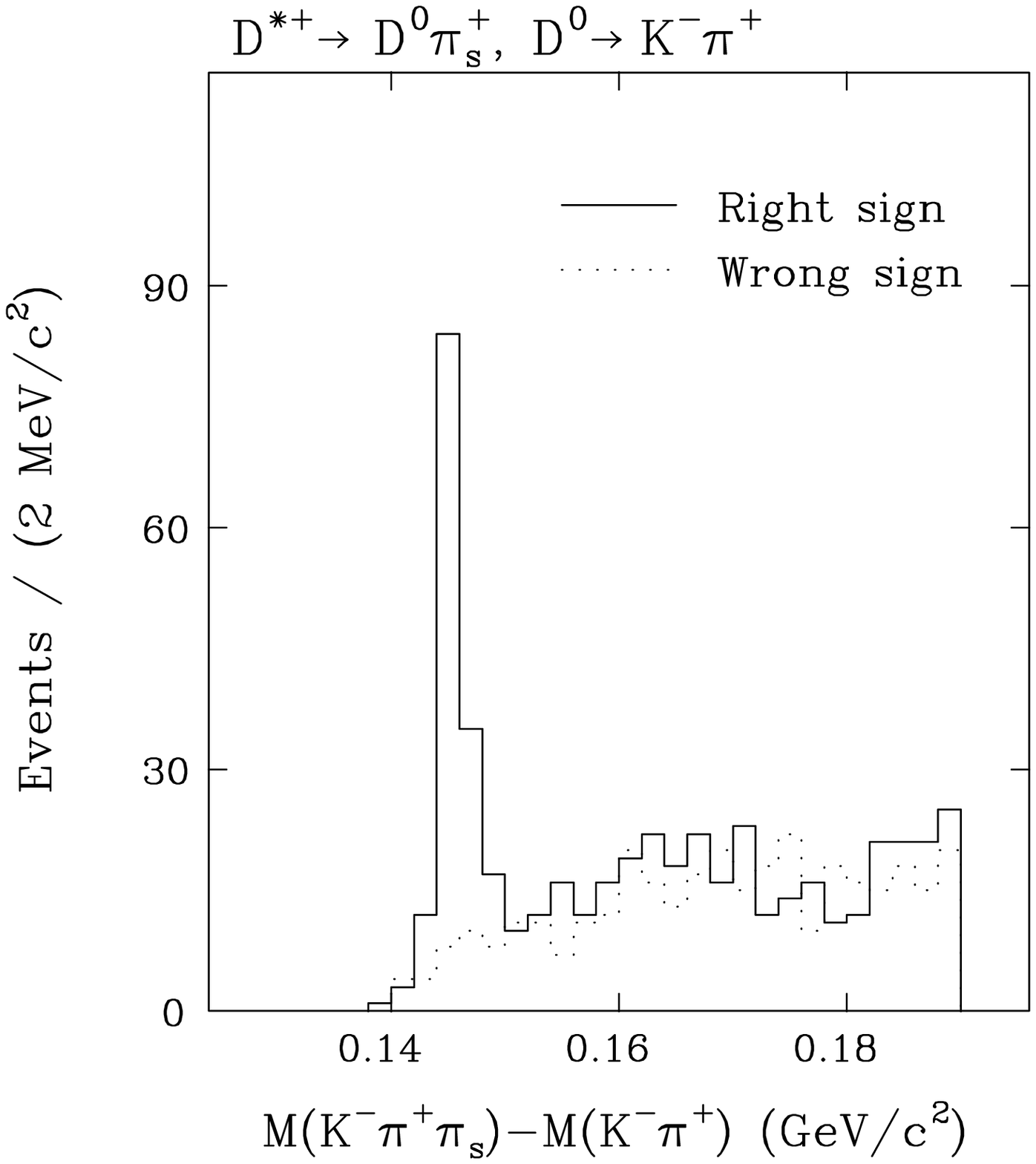}}%
\makebox[0in][r]{\raisebox{3.2in}{\em b) \hspace{2.1in}}}
%\makebox[0in][r]{\raisebox{0.0in}{\em b) \hspace{3.0in}}}
%\makebox[0in][r]{\raisebox{0.0in}{\em b) $\; \; \; \; \; \; \; \;$}}
%\makebox[2.5in][l]{\em b)}
\vspace{-0.8in}
\begin{center}
{\small Fig. 1: a) The $K^- \pi^+$ invariant mass distribution in the
electron sample.  Events from $D^{*+}$ decay are excluded.
b) The distribution of the mass difference, $\Delta m = m(K^- \pi^+ \pi_s) -
m(K^- \pi^+)$, for $D^{*+} \rightarrow D^0 \pi^+_s,
D^0 \rightarrow K^- \pi^+$ candidates.}
\end{center}
%\caption{\small a) The $K^- \pi^+$ invariant mass distribution in the
%electron sample.  Events from $D^{*+}$ decay are excluded and
%$L_{XY} > \sigma(L_XY)$ is required.
%b) The distribution of the mass difference, $\Delta m = m(K^- \pi^+ \pi_s) -
%m(K^- \pi^+)$, for $D^0 \rightarrow K^- \pi^+$ candidates.}
\label{bfig1}
\end{figure}

We now turn to a new, preliminary measurement of the charged and neutral
$B$ meson lifetimes using semileptonic decays~\cite{cdf_blife}, as have been
previously done
by the LEP
experiments~\cite{blife_LEP}.  Partially
reconstructed semileptonic decays of $B$ mesons, namely a lepton in
association with a charm $D^{0}$ or $D^{*+}$ meson will provide {\it nearly
orthogonal} samples of charged and neutral $B$ mesons and thus enable a
determination of their individual lifetimes.
\begin{figure}[htb]
%\hspace{0.05in}
%\parbox[b]{3.2in}{\epsfxsize=3.2in\epsfbox{V615_MS_PHIPI.PS}}%
\parbox[b]{3.0in}{\epsfxsize=3.0in\epsfbox{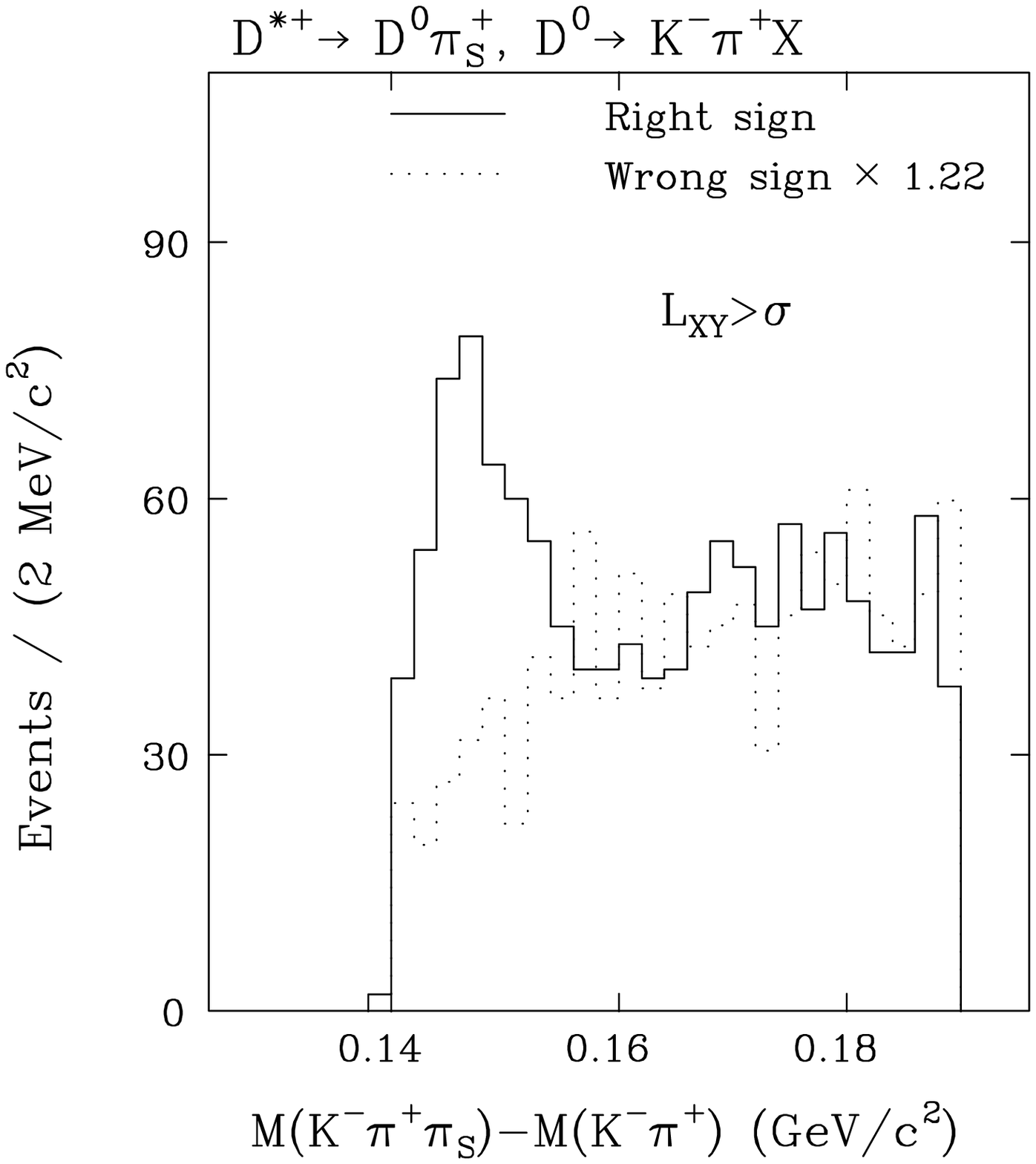}}%
\makebox[0in][r]{\raisebox{3.2in}{\em  a) \hspace{2.1in}}}
%\makebox[0in][r]{\raisebox{0.0in}{\em  a) \hspace{3.0in}}}
%\makebox[0in][r]{\raisebox{0.0in}{\em  a) $\; \; \; \; \; \; \; \;$}}
%\makebox[2.5in][l]{\em a)}
\hfill
%\parbox[b]{3.2in}{\epsfxsize=3.2in\epsfbox{V615_BS_LIFE_LOG.PS}}%
\parbox[b]{3.0in}{\epsfxsize=3.0in\epsfbox{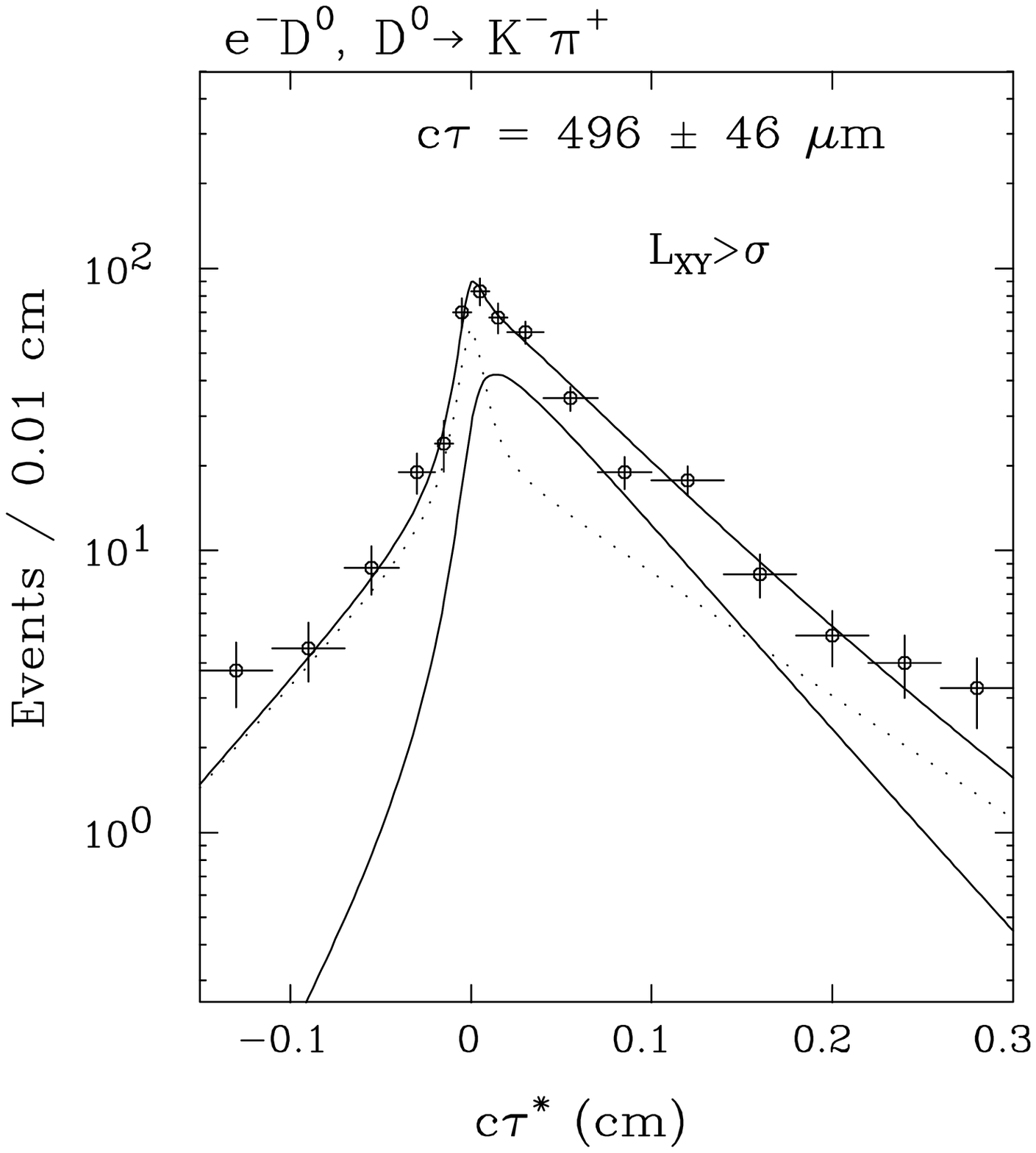}}%
%\parbox[b]{3.0in}{\epsfxsize=3.0in\epsfbox{D0CTAU.PS}}%
\makebox[0in][r]{\raisebox{3.2in}{\em b) \hspace{2.1in}}}
%\makebox[0in][r]{\raisebox{0.0in}{\em b) \hspace{3.0in}}}
%\makebox[0in][r]{\raisebox{0.0in}{\em b) $\; \; \; \; \; \; \; \;$}}
%\makebox[2.5in][l]{\em b)}
\vspace{-0.7in}
\begin{center}
{\small Fig. 2: a) The mass difference distribution for
$D^0 \rightarrow K^- \pi^+ \pi^0$ candidates.
b) The combined, signal, and background
$c\tau^*$ distributions in the $D^0$ signal sample.}
\end{center}
\label{bfig2}
\end{figure}

The present measurement uses only the single electron sample and makes
electron identification cuts which have been described
previously~\cite{electron}.  After applying these cuts, approximately
400,000 electron candidates remain.
The next step is the reconstruction of the $D^{0}$ meson in the decay mode
$D^0 \rightarrow K^- \pi^+$.  This is done by using all oppositely charged
track pairs, assigning kaon and pion masses, and considering only tracks
within a cone of $\Delta R = \sqrt{(\Delta \eta)^2 + (\Delta \phi)^2} < 0.7$
around the electron.  In addition, the momentum of the kaon and pion are
required to satisfy $p(K) > 1.0$ GeV/$c$, $p(\pi) > 0.5$ GeV/$c$ and all
three tracks, the electron, kaon, and pion candidates must be well
reconstructed
within the SVX detector.  Three sources of $D^0$ mesons are considered in this
analysis: 1) $B^- \rightarrow e^- \overline{\nu} D^0 X,
D^0 \rightarrow K^- \pi^+$, where the $D^0$ is {\it not} from $D^{*+}
\rightarrow D^0 \pi^+_s$ (``$D^{0}$ sample''); 2) $\overline{B}^0 \rightarrow
e^-
\overline{\nu} D^{*+} X, D^{*+} \rightarrow D^0 \pi^+_s,
D^0 \rightarrow K^- \pi^+$ (``$D^{*+}$ sample''); and 3) $\overline{B}^0
\rightarrow e^-
\overline{\nu} D^{*+} X, D^{*+} \rightarrow D^0 \pi^+_s, D^0 \rightarrow
K^- \pi^+ \pi^0$ (``satellite sample'') .
The decay length of the $D^0$
in the plane transverse to the colliding beams, $L_{XY}$, must satisfy
$L_{XY} > \sigma(L_{XY})$, where $\sigma(L_{XY})$ is the calculated error on
the transverse decay length, for the $D^{0}$ and satellite samples.
Figure~1a shows the resulting $K^-\pi^+$ mass distribution,
containing $389 \pm 31$ events in the $D^0$ peak, for the $D^{0}$ sample.
Figure~1b gives the mass difference distribution for the
$D^{*+}$ sample, and Fig.~2a gives the $\Delta m$ distribution for
the satellite sample where a cut is made at $\Delta m < 0.155$ GeV/$c^2$.
These figures show the production of $e-D^0$ combinations in the expected
(``right sign'') charge combinations and little evidence above combinatoric
background in the ``wrong sign'' combinations.

	The electron and $D^0$ tracks are then intersected to determine the
$B$ decay vertex position and decay length from the primary vertex.  Since
the $B$ is only partially reconstructed, the $e-D^0$ system transverse
momentum can be used to determine a ``pseudo-$c\tau$'' value
$c\tau^* = L_B m_B/p_{T}(e + D^0) = c\tau/K$,
where $K$ is a momentum correction factor determined from Monte Carlo.  It is
determined separately for each $D^0$ signal sample.

	The signal $c \tau^*$ distributions are fit with an exponential
lifetime term convoluted with a Gaussian resolution function and the
momentum correction distribution.
%\begin{equation}
%F_{SIG}(c\tau^*) = EXP(-c\tau^*;c\tau/K) \otimes G(c\tau^*;\sigma)
%\otimes H(K).
%\end{equation}
The lifetime of the background under the signal peak is determined from the
wrong sign and signal sideband distributions and is modeled by
a Gaussian resolution function plus two exponential tails.
%\begin{equation}
%F_{BG} = (1 - f_{-} - f_{+}) G(c\tau^{*};\sigma) +
%f_{+}EXP(-c\tau^*;\lambda_+)+
%f_{-}EXP(+c\tau^*;\lambda_{-}).
%\end{equation}
Figure~2b shows the result of the lifetime fit for the $D^{0}$
signal sample.  The fit quality and results for the $D^{*+}$ and
satellite samples are similar.
The fraction of $B^{-}$ and $\overline{B^{0}}$ contributing to each of
the $D^{0}, D^{*+}$, and satellite samples is  determined and includes the
effects of cross-talk due to:
1) the $\pi^{+}_{s}$ reconstruction efficiency, $\epsilon(\pi^{+}_{s})
= 0.93 \pm 0.21$.  A missed spectator pion from $D^{*+}$ decay can cause a
$D^{0}$ to be associated with $B^-$ rather than $\overline{B}^0$;
2) the $D^{**}$ fraction, $f^{**} = {\rm BR}(\overline{B}
\rightarrow l^- \overline{\nu} D^{**})/{\rm BR}(\overline{B}
\rightarrow l^- \overline{\nu} X) = 0.36 \pm 0.12$~\cite{CLEO};
3) the fraction of $D^{**}$ decaying to $D^{*}$, from the QQ Monte Carlo
is found to be
${\rm BR}(D^{**} \rightarrow D^* \pi)/
({\rm BR}(D^{**} \rightarrow D^* \pi) + {\rm BR}(D^{**} \rightarrow D \pi))
= 0.78$; and 4)
the charged-to-neutral lifetime ratio can affect the event mixture,
${\rm BR}(B^- \rightarrow l^- \overline{\nu} X)/{\rm BR}(\overline{B^{0}}
\rightarrow l^- \overline{\nu} X) = \tau(B^-)/\tau(\overline{B^0})$.
In spite of these effects, we find that the $D^{0}$ and $D^{*+}$ signals
provide {\it nearly orthogonal} samples of $B^-$ and $\overline{B^0}$ mesons.
A combined likelihood function is used to simultaneously fit the
signal samples for the $B^-$ and $\overline{B^0}$ meson lifetimes.
Variations in the sample composition due to the above effects are included in
the systematic uncertainty.  The results are:
\begin{displaymath}
\begin{array}[b]{rll}
\tau^{+}_{semi} & = & 1.63\pm 0.20 {\rm (stat)} ^{+0.15}_{-0.16} {\rm (syst)
ps} \\
\tau^{0}_{semi} & = & 1.62\pm 0.16 {\rm (stat)} ^{+0.14}_{-0.15} {\rm (syst)
ps} \\
\tau^{+}_{semi}/\tau^{0}_{semi} & = & 1.01\pm 0.19 {\rm (stat)}\pm 0.17
{\rm (syst)}
\end{array}
\end{displaymath}
Tables 1 and 2 show a comparison of the latest~\cite{hessing} $\tau^+$,
$\tau^0$, and $\tau^+/\tau^0$ values from CDF, LEP, and CLEO.  Averaging
asymmetric errors and computing a weighted average, we find that the  error on
the world value for $\tau^+$ and $\tau^0$ is at 5\% and the lifetime ratio has
an uncertainty of 7\%.  Clearly, with some additional statistics and work on
systematic errors, non-spectator contributions to $B$ meson decay will soon be
tested.
\begin{table}[htb]
\begin{center}
{\small Table 1: Comparison of charged and neutral $B$ meson lifetime
measurements.}
\begin{tabular}{|c||c|c||c|} \hline
Experiment & $\tau^{+}$ (ps) & $\tau^{0}$ (ps) &
Reference  \\ \hline
Delphi& $1.30 \pm 0.35$ & $1.17 \pm 0.31$   & Z. Phys. C57, 181 (1993) \\
Delphi& $1.72 \pm 0.10$ & $1.68 \pm 0.21$   & DELPHI 94-97 PHYS 414 \\
Opal  & $1.53 \pm 0.18$ & $1.62 \pm 0.14$   & OPAL Note PN149 \\
Aleph & $1.47 \pm 0.25$ & $1.52 \pm 0.21$   & Phys. Lett. B307, 194 (1993) \\
Aleph & $1.30 \pm 0.23$ & $1.17 \pm 0.22$   & ALEPH Note 94-100 \\
\hline
CDF   & $1.61 \pm 0.17$ & $1.57 \pm 0.20$   & PRL 72, 3456 (1994) \\
CDF   & $1.63 \pm 0.26$ & $1.62 \pm 0.22$   & CDF Note 2598 \\
\hline \hline
World Ave. & $1.60 \pm 0.07$  & $1.53 \pm 0.08$  & DPF 1994 \\
\hline
\end{tabular}
\end{center}
\label{cncomp}
\end{table}
\begin{table}[htb]
\begin{center}
{\small Table 2: Comparison of charged-to-neutral $B$ meson lifetime ratio
results.}
\begin{tabular}{|c||c||c|} \hline
Experiment & $\tau^{+}/\tau^{0}$ & Reference \\ \hline
Delphi& $1.11 \pm 0.46$ & Z. Phys. C57, 181 (1993) \\
Delphi& $1.02 \pm 0.16$ & DELPHI 94-97 PHYS 414 \\
Opal  & $0.94 \pm 0.14$ & OPAL Note PN149 \\
Aleph & $0.96 \pm 0.23$ & Phys. Lett. B307, 194 (1993) \\
Cleo  & $0.93 \pm 0.22$ & CLNS 94/1286~\cite{saulnier} \\
\hline
CDF   & $1.02 \pm 0.17$ & PRL 72, 3456 (1994) \\
CDF   & $1.01 \pm 0.25$ & CDF Note 2598 \\
\hline \hline
World Ave. & $0.98 \pm 0.07$ & DPF 1994 \\
\hline
\end{tabular}
\end{center}
\label{ratcomp}
\end{table}
\begin{table}[hbt]
\begin{center}
{\small Table 3: $B_{s}$ meson lifetime results from CDF and LEP.}
\begin{tabular}{|c||c||c|} \hline
Experiment & $\tau_{s}$ (ps) & Reference \\ \hline
Delphi & $1.42 \pm 0.24$ & Z. Phys. C61, 407 (1994) \\
Opal   & $1.33 \pm 0.24$ & OPAL Note PN113  \\
Aleph & $1.75 \pm 0.36$  & ALEPH Note 94-044   \\
Aleph & $1.92 \pm 0.40$  & Phys. Lett. B322, 275 (1994) \\
\hline
CDF   & $1.42 \pm 0.27$  & CDF Note 2472 \\
CDF   & $1.74 \pm 0.75$  & CDF Note 2515 \\
\hline \hline
World Ave. & $1.49 \pm 0.12$ & DPF 1994  \\
\hline
\end{tabular}
\end{center}
\label{ratbs}
\end{table}
\begin{figure}[thb]
%\hspace{0.05in}
%\parbox[b]{3.2in}{\epsfxsize=3.2in\epsfbox{V615_MS_PHIPI.PS}}%
\parbox[b]{2.8in}{\epsfxsize=2.8in\epsfbox{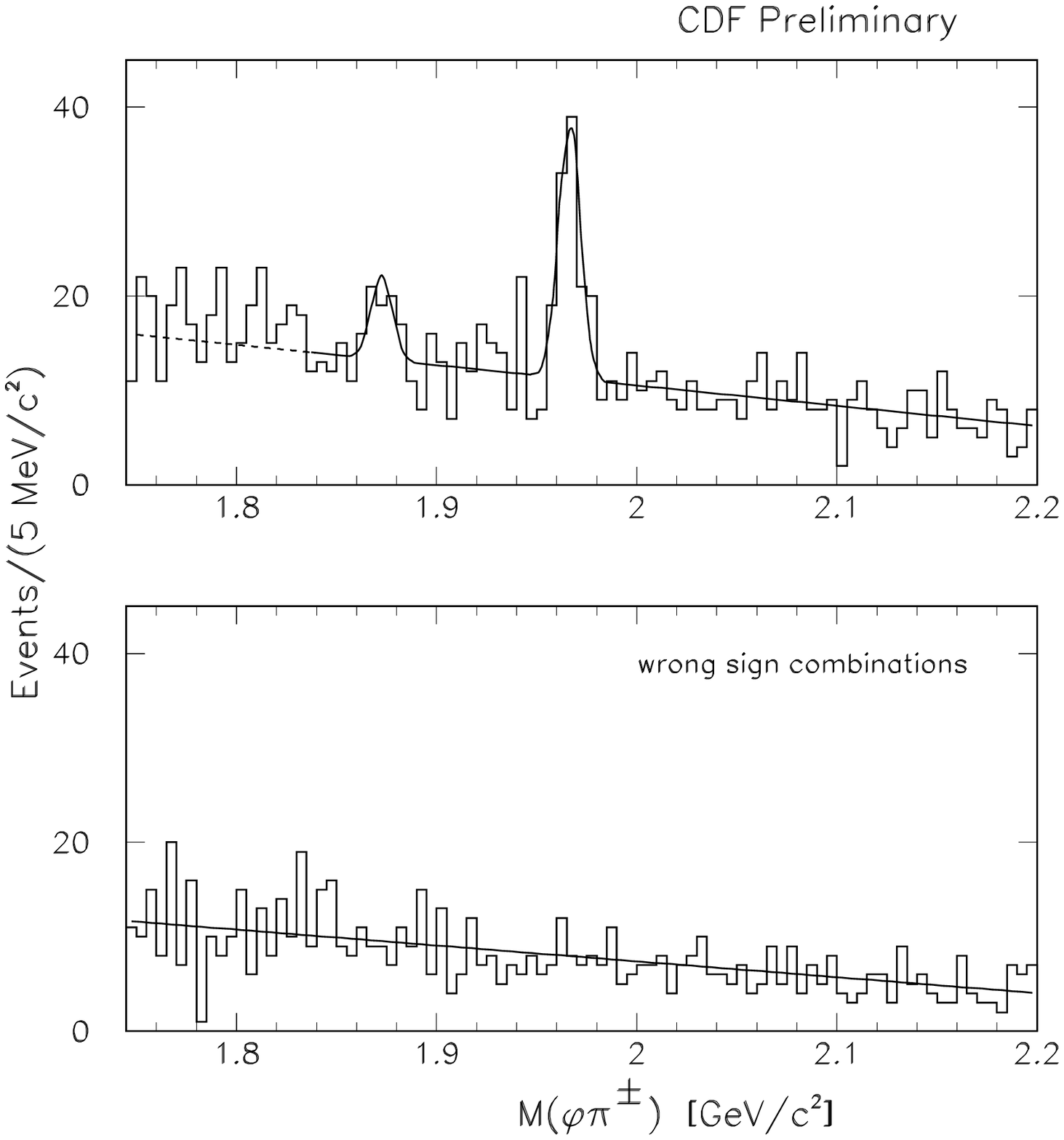}}%
\makebox[0in][r]{\raisebox{3.2in}{\em  a) \hspace{2.3in}}}
%\makebox[0in][r]{\raisebox{0.0in}{\em  a) \hspace{3.0in}}}
%\makebox[0in][r]{\raisebox{0.0in}{\em  a) $\; \; \; \; \; \; \; \;$}}
%\makebox[2.5in][l]{\em a)}
\hfill
%\parbox[b]{3.2in}{\epsfxsize=3.2in\epsfbox{V615_BS_LIFE_LOG.PS}}%
\parbox[b]{2.8in}{\epsfxsize=2.8in\epsfbox{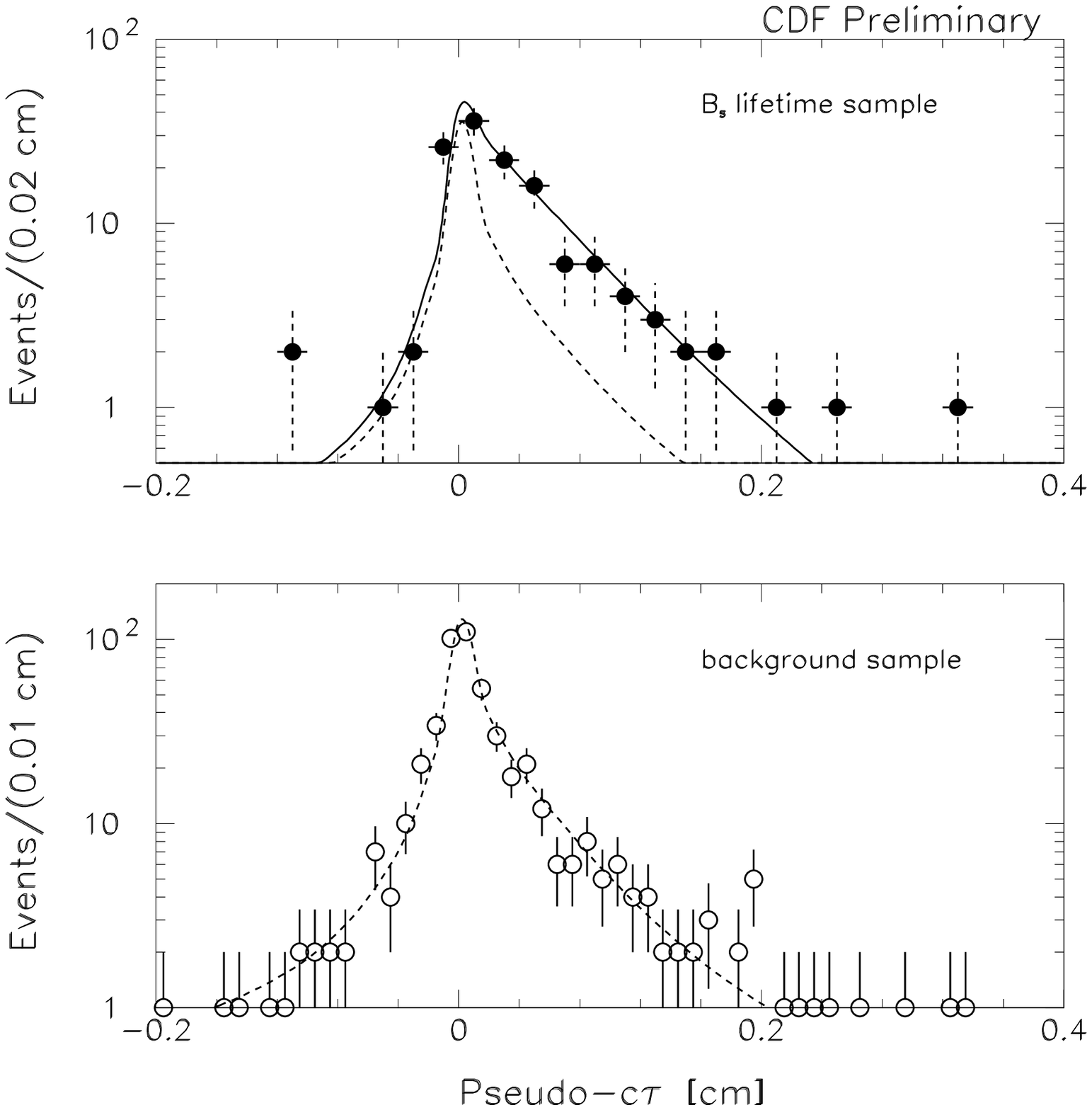}}%
\makebox[0in][r]{\raisebox{3.2in}{\em b) \hspace{2.3in}}}
%\makebox[0in][r]{\raisebox{0.0in}{\em b) \hspace{3.0in}}}
%\makebox[0in][r]{\raisebox{0.0in}{\em b) $\; \; \; \; \; \; \; \;$}}
%\makebox[2.5in][l]{\em b)}
\vspace{-0.6in}
\begin{center}
{\small Fig. 3: a) The $\phi\pi^{-}$ mass distribution for
``right sign'' and ``wrong sign''  lepton-$D_s$ combinations.
b) Pseudo-$c\tau$ distribution of the $l^{-}D^{+}_{s}$ signal
sample showing the lifetime fits of the combined (signal plus background) and
background distributions separately.}
\end{center}
\label{bsfig}
\end{figure}
\section{$B_s$ Meson Lifetime}
A similar technique has been used to measure the $B_s$ meson lifetime using
semileptonic $B_s \rightarrow l \overline{\nu} D_s, D_s \rightarrow \phi \pi,
\phi \rightarrow K^+ K^-$ decays~\cite{cdf_bslife1}.
Figure~3a shows the $K^+ K^- \pi^+$ invariant mass spectrum after
all cuts for the combined electron and muon samples.  Some $76 \pm 8$ events
are found in the right-sign mass peak and a hint of the Cabbibo suppressed
$D^+ \rightarrow \phi \pi^+$ decay is seen.  Following the same procedure as
above, the $B_s$ lifetime from semileptonic $B_s \rightarrow
l \overline{\nu} D_s$ is measured to be (Fig.~3b):
\begin{displaymath}
\begin{array}[b]{rll}
\tau_{s}^{semi} & = & 1.42 ^{+0.27}_{-0.23} {\rm (stat)} \pm 0.11 {\rm (syst)
ps} \\
\end{array}
\end{displaymath}
Finally, there is a new low-statistics measurement of the $B_s$ lifetime using
fully reconstructed $B_s \rightarrow J/\psi \phi, J/\psi
\rightarrow \mu^+ \mu^-,
\phi \rightarrow K^+ K^-$ decays~\cite{cdf_bslife2}.  At least two of the four
daughter tracks are
required to be reconstructed in the SVX.  Based on a sample of 10 events, the
$B_s$ lifetime using exclusive $B_s \rightarrow J/\psi \phi$ is measured to be:
\begin{equation}
\tau_s^{exc} = 1.74 ^{+0.90}_{-0.60} ({\rm stat}) \pm 0.07 ({\rm syst}) {\rm
ps}
\end{equation}
Table 3 shows a comparison of the latest $B_s$ lifetime measurements at CDF
and LEP~\cite{karlen}.
Calculated similar to above, the world average value has only an 8\%
uncertainty.
\section{Conclusions}
	The first measurements of the $B_u$, $B_d$, $B_s$ meson lifetimes and
the $B_u/B_d$ lifetime ratio by CDF are comparable to the latest results
from other  experiments.  With the installation of a rad-hard SVX and possibly
x5 more data from the present Run Ib, the prospects for precision measurements
of the $B$ meson lifetimes in both inclusive and exclusive modes in the near
future is very promising.
\bibliographystyle{unsrt}

\begin{thebibliography}{99.}
\bibitem{CDF} F. Abe et al. (CDF Collaboration),
{\it Nucl. Instrum. Methods Phys. Res., Sect. A}
{\bf 271} (1988) 387, and references therein.
\bibitem{svx} D. Amidei et al., preprint FERMILAB-PUB-94/024-E,
submitted to {\it Nucl. Instrum. Methods Phys. Res.}
\bibitem{thlife} M. B. Voloshin and M. A. Shifman, {\it Sov. Phys. JETP}
{\bf 64}, 698 (1986), and more recently V. Chernyak, preprint BudkerINP 94-69.
\bibitem{charge}  Throughout this paper, references to a specific charge state
imply the charge-conjugate state as well.
\bibitem{cdf_exlife} F. Abe et al., {\it Phys. Rev. Lett.} {\bf 72} 3456
(1994).
\bibitem{cdf_blife} CDF Coll., {\it CDF internal note 2598}.
\bibitem{blife_LEP} ALEPH Coll., {\it Phys. Lett.} {\bf B307}, 194 (1993);
OPAL Coll., {\it OPAL internal note PN149}; DELPHI Coll., {\it Z. Phys.}
{\bf C57}, 181 (1993).
\bibitem{electron} F. Abe et al., {\it Phys. Rev. Lett.} {\bf 71} 500 (1993).
\bibitem{CLEO} CLEO Coll., {\it Phys. Rev.} {\bf D43}, 651 (1991).
\bibitem{hessing} T. Hessing (DELPHI), private communication.
\bibitem{saulnier} M. Saulnier (CLEO), private communication.
\bibitem{cdf_bslife1} CDF Coll., {\it CDF internal note 2472}.
\bibitem{cdf_bslife2} CDF Coll., {\it CDF internal note 2515}.
\bibitem{karlen} D. Karlen (OPAL), private communication.
\end{thebibliography}

\end{document}